\documentclass[aps,twocolumn,showpacs,prl,preprintnumbers,amsmath,amssymb]{revtex4}

\usepackage[dvipdfmx]{graphicx}
\usepackage{dcolumn}
\usepackage{bm}

\usepackage[dvipdfmx]{color}
\usepackage{ulem}

\begin{document}

\preprint{APS/123-QED}

\title{Fractional Spin Fluctuation as a Precursor of Quantum Spin Liquids:\\
Majorana Dynamical Mean-Field Study for the Kitaev Model
}

\author{Junki Yoshitake$^1$, Joji Nasu$^2$, and Yukitoshi Motome$^1$}
\affiliation{%
$^1$Department of Applied Physics, University of Tokyo, Bunkyo, Tokyo 113-8656, Japan\\
$^2$Department of Physics, Tokyo Institute of Technology, Meguro, Tokyo 152-8551, Japan
}%




\date{\today}

\begin{abstract}
Experimental identification of quantum spin liquids remains a challenge, as the pristine nature is to be seen in asymptotically low temperatures. 
We here theoretically show that the precursor of quantum spin liquids appears in the spin dynamics in the paramagnetic state over a wide temperature range.
Using the cluster dynamical mean-field theory and the continuous-time quantum Monte Carlo method, which are newly developed in the Majorana fermion representation, we calculate the dynamical spin structure factor, relaxation rate in nuclear magnetic resonance, and magnetic susceptibility for the honeycomb Kitaev model whose ground state is a canonical example of the quantum spin liquid. 
We find that dynamical spin correlations show peculiar temperature and frequency dependence even below the temperature where static correlations saturate. 
The results provide the experimentally-accessible symptoms of the fluctuating fractionalized spins evincing the quantum spin liquids.
\end{abstract}

\maketitle

The quantum spin liquid (QSL) has attracted much attention for decades, as a new state of matter in insulating magnets stabilized by quantum fluctuations~\cite{Anderson1973}. 
Although several candidate materials have been studied, experimental identification of QSLs still remains a challenge in modern condensed matter physics~\cite{Balents2010,Lacroix2011}. 
This is mainly because of the absence of conventional order parameters: it is hard to prove the lack of any symmetry breaking down to the lowest temperature ($T$). 
Many attempts were made also on the low-$T$ behavior of thermodynamic quantities, for instance, the specific heat~\cite{Helton2007,Okamoto2007,Yamashita2008}, which reflect the low-energy excitations specific to QSLs. 
All these efforts are nonetheless extremely difficult, as the asymptotically low-$T$ physics might be sensitively affected by extrinsic factors, such as impurities and subordinate interactions. 

On the other hand, QSLs are established in several theoretical models. 
Among them, the Kitaev model provides a canonical example of exact QSLs with fractional excitations in the ground state~\cite{Kitaev2006}.
The model is believed to describe the anisotropic exchange interactions realized in insulating magnets with strong spin-orbit coupling, such as Ir oxides~\cite{Jackeli2009}. 
This has stimulated a new trend of exploration of QSLs in real materials~\cite{Singh2010,Singh2012,Plumb2014,Kubota2015}. 
Recently, several experimental efforts have been made on the identification of the fractional excitations in the paramagnetic state above the N\'eel temperature as a precursor to QSLs~\cite{Gretarsson2013,Sears2015,Sandilands2015,Banerjee2016}. 
Indeed, such a signature in a wide $T$ range was theoretically predicted for thermodynamic quantities~\cite{Nasu2015}. 
However, the signature of fractionalization is most clearly visible in the dynamics, for which theoretical studies were limited to the ground state~\cite{Knolle2014,Knolle2014b}. 
Thus, the `missing link' between theory and experiment exists
in the dynamical properties in the experimentally-accessible $T$ range.
This is, however, a theoretical challenge as it requires to handle both quantum and thermal fluctuations simultaneously. 

In this Letter, we present numerical results on the dynamical properties of the Kitaev model at finite $T$.
To take into account quantum and thermal fluctuations on an equal footing, we develop the cluster dynamical mean-field theory (CDMFT) and the continuous-time quantum Monte Carlo method (CTQMC) in the Majorana fermion representation of this quantum spin model. 
We calculate the experimentally measurable quantities: the dynamical spin structure factor, $S(\textbf{q},\omega),$ which is measured in the neutron scattering experiment, the relaxation rate in nuclear magnetic resonance (NMR), $1/T_1$, and the magnetic susceptibility $\chi$, for both ferromagnetic (FM) and antiferromagnetic (AFM) cases.
We show that the dynamical spin fluctuations in the paramagnetic state are strongly influenced by the thermal fractionalization of quantum spins.
$S(\textbf{q},\omega)$ exhibits the growth of inelastic and quasi-elastic responses at very different $T$ scales.
Also, $1/T_1$ begins to increase below the temperature where the static spin correlations saturate, and shows a peak at very low $T$, despite the suppression of $\chi$ from the Curie-Weiss behavior.
These unconventional features will provide a smoking gun for fractionalized spins in the Kitaev-type QSLs.

\begin{figure}[t]
    \includegraphics[width=\columnwidth,clip]{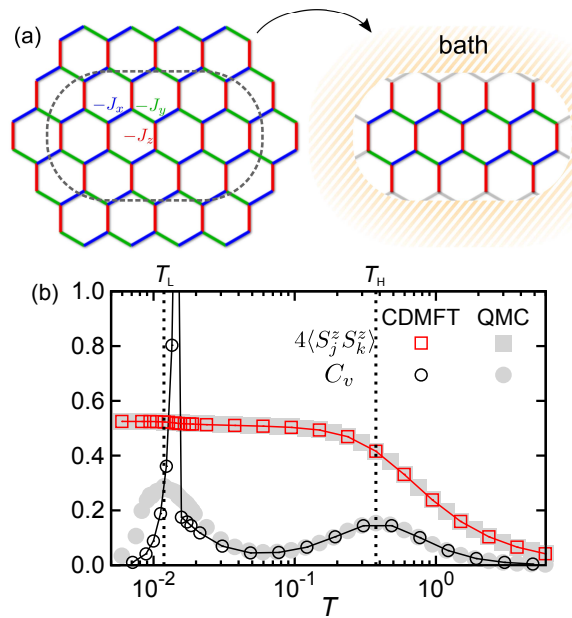}
    \caption{ \label{fig:fig1} 
    (a) Schematic picture of the Kitaev model on the honeycomb lattice  [Eq.~(\ref{eq:Hamiltonian0})] and the mapping to a 26-sites cluster used in the Majorana CDMFT.
    (b) The specific heat $C_v$ and equal-time spin correlations for NN sites, $\langle S_j^z S_k^z \rangle$, obtained by the Majorana CDMFT for the isotropic FM case.
    QMC data in Ref.~\cite{Nasu2015} are plotted by gray symbols for comparison.
    }
\end{figure}

 We consider the Kitaev model on a honeycomb lattice, whose Hamiltonian is given by~\cite{Kitaev2006}
\begin{align}
\mathcal{H} = 
-J_x \sum_{{\langle j,k\rangle}_x} S_j^x S_k^x
-J_y \sum_{{\langle j,k\rangle}_y} S_j^y S_k^y
- J_z \sum_{{\langle j,k\rangle}_z} S_j^z S_k^z,
\label{eq:Hamiltonian0}
\end{align}
where $S_{j}^p$ is the $p(=x,y,z)$ component of the $S=1/2$ spin at site $j$.
The sum of ${\langle j,k \rangle}_p$
is taken for the nearest-neighbor (NN) sites on three inequivalent bonds of the honeycomb lattice, as indicated in Fig.~\ref{fig:fig1}(a).

The exact solution for the ground state of the model (\ref{eq:Hamiltonian0}) is obtained by introducing Majorana fermions~\cite{Kitaev2006}.
A formulation, which is suitable for the following numerical calculations at finite $T$, is obtained by
applying the Jordan-Wigner transformation to the one-dimensional chains composed of the $J_x$ and $J_y$ bonds and introducing 
two types of Majorana fermions $c_j$ and $\bar{c}_j$~\cite{Chen2007,Feng2007,Chen2008}. Then, the Hamiltonian in Eq.~(\ref{eq:Hamiltonian0}) is rewritten as
\begin{align}
\mathcal{H}=
i\frac{J_x}{4} \sum_{(j,k)_x} c_j c_k
- i\frac{J_y}{4} \sum_{(j,k)_y} c_j c_k
- i\frac{J_z}{4} \sum_{(j,k)_z} \eta_r c_j c_k,
\label{eq:Hamiltonian1}
\end{align}
where $(j,k)_p$ is the NN pair satisfying $j<k$ on the $p$ bond.
Here, $\eta_r = i\bar{c}_j \bar{c}_k$ is defined on each $z$ bond ($r$ is the bond index); 
$\eta_r$ is a $Z_2$ variable taking $\pm 1$, as $\eta_r^2=1$ and it commutes with the Hamiltonian as well as all the other $\eta_{r'}$.
The ground state is given by all $\eta_r=1$, dictating a QSL with gapless or gapful excitations depending on the anisotropy in the coupling constants~\cite{Kitaev2006}.

At finite $T$, however, the configuration of $\{ \eta \}$ is disturbed by thermal fluctuations. 
Hence, the model in Eq.~(\ref{eq:Hamiltonian1}) describes itinerant Majorana fermions coupled to thermally-fluctuating classical variables $\eta_r$, which can be regarded as a variant of the double-exchange model.
This allows one to utilize theoretical tools developed for fermion systems, such as the quantum Monte Carlo method~\cite{Nasu2014}. 
In this study, we construct the CDMFT~\cite{Kotliar2001} for this Majorana fermion problem. 
By following the formulation for the double-exchange model~\cite{Furukawa1994} and using the path-integral representation for Majorana fermions~\cite{Nilsson2013}, the effective action for a cluster embedded in a bath [see Fig.~\ref{fig:fig1}(a)] is given by
\begin{align} 
\mathcal{S}_{\text{eff}}^{\{\eta\}} =
& - T{\sum_{j,k,n \geq 0
}} \chi_{j,-\omega_n} (\mathcal{G}_{0}(i\omega_n))_{j,k}^{-1} \chi_{k,\omega_n} \nonumber\\
&+i \frac{J_z}{2} T{\sum_{\langle j,k\rangle_{z},n}} \eta_{r}\chi_{j,-\omega_n}\chi_{k,\omega_n},
\label{eq:EffAction}
\end{align}
where $\chi_{j,\omega_n}$ is the Grassmann number corresponding to the Majorana operator $c_j$, and $\mathcal{G}_{0}$ represents the Weiss function including the effect of bath.
For a given configuration of $\{\eta\}$, the impurity problem is exactly solvable and Green's function is obtained as $(G^{\{\eta\}}(i\omega_n))^{-1} = (\mathcal{G}_{0}(i\omega_n))^{-1} - h^{\{\eta\}}$, where $h^{\{\eta\}}$ is the matrix representation of the second term in Eq.~(\ref{eq:EffAction}). 
Local Green's function is also exactly calculated through $G(i\omega_n) = \text{P}(\{\eta\})G^{\{\eta\}}(i\omega_n) $, where $\text{P}(\{\eta\}) = Z^{\{\eta\}}/\sum_{\{\eta\}}Z^{\{\eta\}}$ with $Z^{\{\eta\}} = e^{-\mathcal{S}_{\text{eff}}^{\{\eta\}}}=\prod_{n \geq 0
} \text{det}[-G^{\{\eta\}}(i\omega_n)]$, as we can compute $G^{\{\eta\}}(i\omega_n)$ and $\text{P}(\{\eta\})$ for all $2^{N_c/2}$ configurations of $\{\eta\}$ in a $N_c$-site cluster~\cite{Udagawa2012}.
The self-consistent equations in CDMFT are given as $\Sigma(i\omega_n) = (\mathcal{G}_{0}(i\omega_n))^{-1} - (G(i\omega_n))^{-1}$ and $(\mathcal{G}_{0}(i\omega_n))^{-1} =
\left(\frac{1}{N}\sum_{\mathbf{k}} 
[i\omega_n - 2\mathcal{H}_0(\mathbf{k}) - \Sigma(i\omega_n)]^{-1} \right)^{-1}
 + \Sigma(i\omega_n)$, where $\Sigma$ is the self-energy, $N$ is the number of clusters in the whole lattice, and $\mathcal{H}_0(\mathbf{k})$ is the Fourier transform of the first and second terms in Eq.~(\ref{eq:Hamiltonian1}).
 
Thus, the Majorana CDMFT provides the {\it exact} results for thermodynamic quantities of the quantum spin model (\ref{eq:Hamiltonian0}), except for the cluster approximation. 
This is a distinct advantage of the Majorana representation: the original quantum spin representation does not admit the exact enumeration. 
In addition, the cluster approximation works quite well in the current system with extremely short-range spin correlations~\cite{Baskaran2007}, as demonstrated below.
In the following calculations, we take the 26-sites cluster shown in Fig.~\ref{fig:fig1}(a), and consider $60 \times 80$ array of the unit cell~\cite{supp}.
Typically, the CDMFT loop is repeated for ten times until convergence.

A benchmark of the Majorana CDMFT is shown in Fig.~\ref{fig:fig1}(b).
We compare the specific heat and the equal-time NN spin correlations $\langle S_j^p S_k^p \rangle$ obtained by CDMFT with those by QMC in Ref.~\cite{Nasu2015}. 
The data are calculated for the isotropic FM case, $J_x=J_y=J_z=1$ (the sign of $\langle S_j^p S_k^p \rangle$ is reversed for AFM). 
As indicated by two broad peaks in the specific heat in the QMC results, the system exhibits two crossovers at $T_{\rm H} \sim 0.375$ and $T_{\rm L} \sim 0.012$.
The spin correlations grow down to $T\sim T_{\rm H}$, while they saturate below $T_{\rm H}$ and do not show significant changes at $T_{\rm L}$~\cite{Nasu2015}.
These behaviors are excellently reproduced by CDMFT, except for the low-$T$ peak in the specific heat. 
The sharp anomaly in the CDMFT result at $T\simeq 0.014$ is due to a phase transition by ordering of $\langle \eta \rangle$, which is an artifact of the mean-field nature of CDMFT. 
The comparison, however, shows that the CDMFT gives qualitatively correct results in a wide $T$ range above the low-$T$ crossover, i.e., $T \gtrsim  0.015$. 
We note that the quantum spin liquid state at sufficiently low $T$, where all $\eta_r=1$, is also reproducible~\cite{supp}. 
Thus, the present CDMFT enables to calculate the physical properties with sufficient precision in the wide $T$ range except for the vicinity of $T_{\rm L}$. 
In the following, we apply the CDMFT in this qualified $T$ range above $T_{\rm L}$ to the study of spin dynamics, which one cannot compute by QMC. 

In the calculations of dynamical quantities, we need an additional effort beyond the exact enumeration in the Majorana CDMFT.
This is because the calculation of dynamical spin correlations $\langle S^{p}_{j}(\tau)S^{p}_{k} \rangle$ requires the imaginary time evolution of the {$\bar{c}$ variables that compose the conserved quantities $\eta$.
To compute $\langle S^{p}_{j}(\tau)S^{p}_{k} \rangle$, we adopt the CTQMC based on the strong coupling expansion~\cite{Werner2006}.
$\langle S^{z}_{j}(\tau)S^{z}_{k} \rangle$ on an $r_0$ bond is calculated as $\langle S^{z}_{j}(\tau)S^{z}_{k} \rangle = \sum_{\{\eta\}', \eta_{r_0} = \pm 1}\text{P}(\{\eta\}', \eta_{r_0}) \langle S^{z}_{j}(\tau)S^{z}_{k} \rangle^{\{\eta\}' }$ by using the CDMFT solutions; here, $\{\eta\}'$ represents the configurations of $\eta_r$ except for the $r_0$ bond, and $\langle S^{z}_{j}(\tau)S^{z}_{k} \rangle^{\{\eta\}' }$ is calculated for a given $\{\eta\}'$ by CTQMC. 
As the interaction between Majorana fermions lies only on the $r_0$ bond, it is sufficient to solve the two-site impurity problem in CTQMC.
In the CTQMC calculations, we typically run $10^7$ steps with measurement at every 20 steps, after $10^5$ steps of initial relaxation, for each $\langle S^{z}_{j}(\tau)S^{z}_{k} \rangle^{\{\eta\}' }$.
$\langle S^{p}_{j}(\tau)S^{p}_{k} \rangle$ for $p=x, y$ are obtained by taking the lattice coordinate so that $p=z$. 
In the following, we present the results for the isotropic coupling, $J_x=J_y=J_z=J$, where the ground state is a gapless quantum spin liquid~\cite{Kitaev2006}. 
We compute both FM and AFM cases~\cite{note} with setting $|J|=1$ as the energy unit. 
The systematic study of the anisotropic cases will be reported elsewhere.

Using the Majorana CDMFT+CTQMC, we calculate the dynamical spin structure factor, NMR relaxation rate, and magnetic susceptibility.
The dynamical spin structure factor is defined as $S(\textbf{q}, \omega)=1/(3NN_c)\sum_p\sum_{j,k} e^{i\textbf{q}\cdot(\textbf{r}_{j}-\textbf{r}_{k})}S^{p}_{j,k}(\omega)$, where $S^{p}_{j,k}(\omega)$ is obtained by solving $\langle S^{p}_{j} (\tau)S^{p}_{k} \rangle = \int d\omega S^{p}_{j,k}(\omega)e^{-\omega\tau}$ by the maximum entropy method~\cite{Jarrell1996}.
We confirmed the validity of the procedures by the fact that the low-$T$ result with $\langle\eta\rangle\simeq1$ (beyond the quantified $T$ range) reproduces the $T=0$ solution~\cite{Knolle2014}. 
The NMR relaxation rate is obtained by using the relation, $1/T_1 \propto S^{x}_{j,k}(\omega = 0) + S^{y}_{j,k}(\omega = 0)$; we compute the contributions from onsite and NN-site correlations separately, as the hyperfine coupling is unknown. 
The magnetic susceptibility is calculated as $\chi^p = 1/(NN_c) \sum_{j,k}\int d\tau\langle S_{j}^{p}(\tau)S_{k}^{p}\rangle$; $\chi^x=\chi^y=\chi^z=\chi$ for the isotropic coupling. 

\begin{figure}[t]
    \includegraphics[width=\columnwidth,clip]{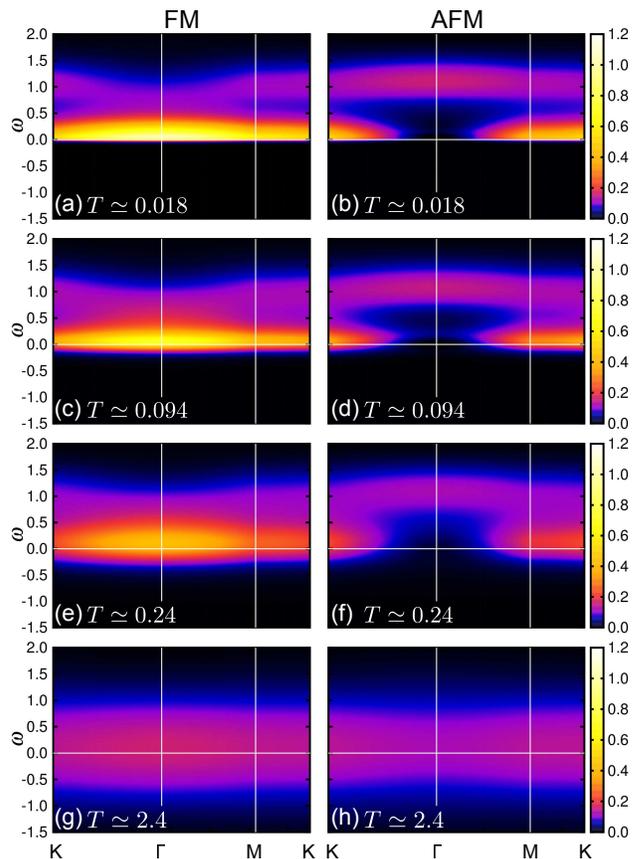}
    \caption{ \label{fig:fig2} 
    The dynamical spin structure factor $S(\mathbf{q}, \omega)$ obtained by the Majorana CDMFT+CTQMC for the (a)(c)(e)(g) FM and (b)(d)(f)(h) AFM cases at (a)(b) $T\simeq0.018$, (c)(d) $T\simeq0.094$, (e)(f) $T\simeq0.24$, and (g)(h) $T\simeq2.4$.
    }
\end{figure}

Figure~\ref{fig:fig2} shows the $T$ dependences of the dynamical spin structure factor $S(\textbf{q},\omega)$ for both FM and AFM cases. 
At high $T>T_{\rm H}$, $S(\textbf{q},\omega)$ shows only a diffusive response at $\omega\sim 0$ with less $\textbf{q}$ dependence in both FM and AFM cases, as shown in Figs.~\ref{fig:fig2}(g) and \ref{fig:fig2}(h). 
The $\textbf{q}$-$\omega$ dependence begins to develop while lowering $T$ below $T_{\rm H}\sim 0.375$; the diffusive weight shifts to a positive $\omega$ region ranging to $\omega\sim J$ below $T_{\rm H}$ [Figs.~\ref{fig:fig2}(e) and \ref{fig:fig2}(f)], and at the same time, the quasi-elastic component at $\omega\sim0$ grows gradually [Figs.~\ref{fig:fig2}(c) and \ref{fig:fig2}(d)]. 
The quasi-elastic response is large around the $\Gamma$ point in the FM case, whereas it is distributed on the Brillouin zone boundary (along the K-M line) in the AFM case. 
With further decreasing $T$, the intensity of the quasi-elastic peak continues to increase while approaching $T_{\rm L}\sim 0.012$, as shown in Figs.~\ref{fig:fig2}(a) and \ref{fig:fig2}(b). 
The low-$T$ behavior converges on the ground state solution, which has a sharp peak at $\omega\sim 0.12J$ together with an incoherent weight at $\omega\sim J$~\cite{Knolle2014}. 

\begin{figure}[t]
    \includegraphics[width=\columnwidth,clip]{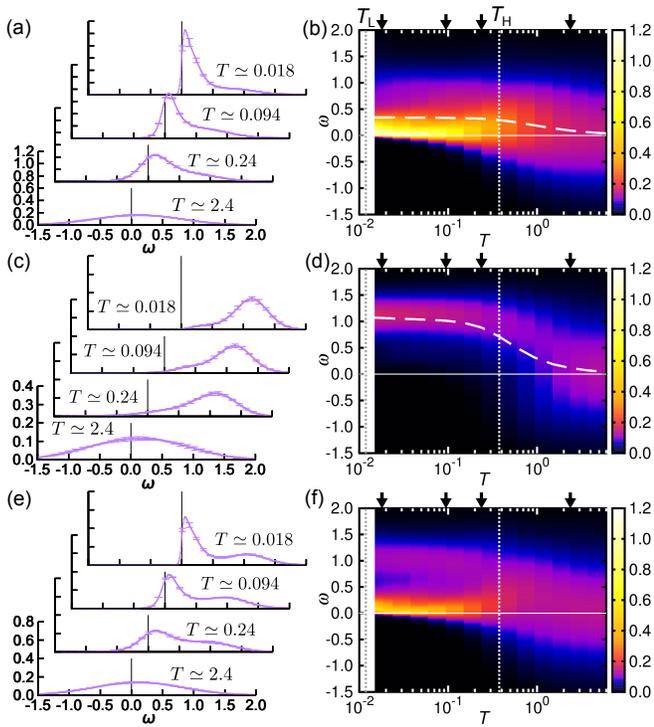}
    \caption{ \label{fig:fig3} 
    $S(\Gamma,\omega)$ for the (a) FM and (c) AFM cases, and (e) $S({\rm{K}},\omega)$ at several $T$.
    (e) is common to the FM and AFM cases.
    The corresponding contour plots in the $T$-$\omega$ plane are shown in (b)(d)(f).
    The arrows indicate the temperatures used for the data in (a)(c)(e).
    The dashed curves represent the average frequency of  $S(\mathbf{q}, \omega)$ (see the text for details).
    In (a)(c)(e), the errorbars are shown for every ten data along the $\omega$ axis. 
    }
\end{figure}

To see these behaviors more clearly, we show the data at the $\Gamma$ and K points, denoted as $S(\Gamma,\omega)$ and $S({\rm K},\omega)$, respectively, in Fig.~\ref{fig:fig3}. Note that $S({\rm K},\omega)$ is the same for the FM and AFM cases by symmetry.
In the FM case, $S(\Gamma,\omega)$ and $S({\rm K},\omega)$ show qualitatively similar $T$-$\omega$ dependence, as shown in Figs.~\ref{fig:fig3}(a)(b) and \ref{fig:fig3}(e)(f); the inelastic response at $\omega\sim J$ appears below $T_{\rm H}$, and the quasi-elastic one at $\omega\sim0$ rapidly grows as approaching $T_{\rm L}$.
On the other hand, in the AFM case, the strong quasi-elastic intensity at low $T$ is absent, while the inelastic response at $\omega\sim J$ arises below $T_{\rm H}$, as in the FM case, as shown in Figs.~\ref{fig:fig3}(c)(d).

Despite the different $\textbf{q}$ dependence reflecting the sign of $J$, $S(\textbf{q},\omega)$ exhibits common characteristic $\omega$-$T$ dependence: the emergence of the inelastic response at $\omega\sim J$ for $T\lesssim T_{\rm H}$, and the rise of the quasi-elastic response as $T\to T_{\rm L}$. 
These peculiar behaviors are regarded as the signatures of the fractionalization of quantum spins into two types of Majorana fermions, itinerant ``matter fermions" and localized ``fluxes", which correspond to $c$ and $\eta$ in Eq.~(\ref{eq:Hamiltonian1}), respectively. 
The previous QMC studies revealed that the matter fermions and fluxes affect the thermodynamics at very different $T$ scales~\cite{Nasu2015,Nasu2014}; the kinetic energy of matter fermions, which is equivalent to the equal-time spin correlations $\langle S_j^{p}S_k^{p} \rangle$ [see Eqs.~(\ref{eq:Hamiltonian0}) and (\ref{eq:Hamiltonian1})], is gained at $T\sim T_{\rm H}$, whereas the fluxes shows a condensation at $T\sim T_{\rm L}$ to the flux-free state with all $\eta_r=1$. 
Our results of $S(\textbf{q},\omega)$ obtained above indicate that the former is closely related to the evolution of the inelastic response at $\omega\sim J$ below $T_{\rm H}$, while the latter to the rise of the quasi-elastic response as approaching $T_{\rm L}$. 

The relation between the inelastic response and the matter fermions is grasped through two sum rules for the dynamical spin structure factor. 
For instance, at the $\Gamma$ point, the sum rules read} $\int S^{p}(\Gamma,\omega) d\omega = \langle S_j^{p}S_k^{p} \rangle + 1/4$ and $\int \omega S^{z}(\Gamma,\omega) d\omega = (J_x \langle S_j^{x}S_k^{x} \rangle + J_y \langle S_j^{y}S_k^{y} \rangle)/2$ (similarly for $p=x,y$), where $\langle S_j^p S_k^p \rangle$ denotes the equal-time NN spin correlation~\cite{Hohenberg1974}. 
As mentioned above, in the Kitaev model, $\langle S_j^{p}S_k^{p} \rangle$ corresponds to the kinetic energy of the matter fermions.
Hence, when we compute the average frequency of  $S^{p}(\Gamma,\omega)$ by the ratio of the two sum rules, $\bar\omega \equiv \int \omega S^{z}(\Gamma,\omega) d\omega / \int S^{z}(\Gamma,\omega) d\omega$, the kinetic energy gain of the matter fermions at $T\sim T_{\rm H}$ results in the shift of $\bar\omega$ from almost zero to a nonzero positive value. 
This is indeed seen in Figs.~\ref{fig:fig3}(b) and \ref{fig:fig3}(d), where $\bar\omega$ is plotted by the dashed curves. 
The difference of the values of low-$T$ $\bar\omega$ between the FM and AFM cases is also accounted for by the opposite sign of the NN contribution in the first sum rule.

On the other hand, below $T\sim T_{\rm L}$, the quasi-elastic response converges on the sharp peak at $\omega\sim 0.12J$ with a flux gap $\Delta\simeq 0.065J$~\cite{Kitaev2006} in the $T=0$ solution~\cite{Knolle2014}. 
As the fluxes are proliferated above $T\sim T_{\rm L}$~\cite{Nasu2015}, the decay of the quasi-elastic response for $T\gtrsim T_{\rm L}$ is considered as a consequence of the thermally excited fluxes. 
Indeed, the flux gap is smeared out in our results above $T\sim T_{\rm L}$. 

\begin{figure}[t]
    \includegraphics[width=1.0\columnwidth,clip]{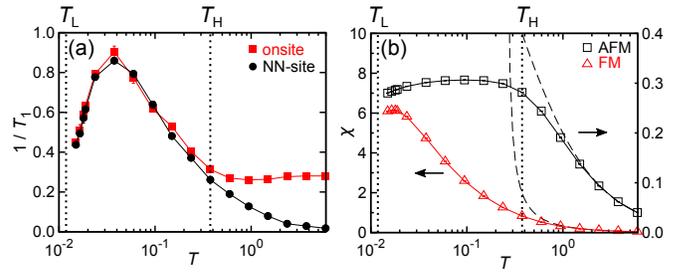}
    \caption{ \label{fig:fig4} 
    $T$ dependences of (a) the NMR relaxation rate $1/T_1$ and (b) the magnetic susceptibility $\chi$. 
    In (b), the dashed curves represent the Curie-Weiss behavior $\chi_{\rm CW}$. 
    }
\end{figure}

We find that the influence of excited fluxes is more clearly visible in the NMR relaxation rate $1/T_1$. 
Figure~\ref{fig:fig4}(a) shows the CDMFT+CTQMC results of $1/T_1$: we plot the value of $S^{x}_{j,k}(\omega = 0) + S^{y}_{j,k}(\omega = 0)$ in the FM case (the NN component changes its sign in the AFM case).
In the high-$T$ region above $T_{\rm H}$, as expected for the conventional paramagnets~\cite{Moriya1956}, the onsite component is nearly $T$ independent, while the NN-site one increases gradually with decreasing $T$, reflecting the growth of equal-time spin correlations $\langle S^{p}_j S^{p}_k \rangle$ in Fig.~\ref{fig:fig1}(b). 
Below $T_{\rm H}$, however, both components increase and show a peak at slightly above $T_{\rm L}$, despite the saturation of equal-time correlations~\cite{note2}.
The pronounced peak is regarded as the consequence of thermally excited fluxes above $T_{\rm L}$, as the suppression of $1/T_1$ for $T\lesssim T_{\rm L}$ is due to the formation of the flux gap in the low-$T$ limit~\cite{Knolle2014}. 
The unexpected behavior below $T_{\rm H}$ is also seen in comparison with the magnetic susceptibility $\chi$ in Fig.~\ref{fig:fig4}(b); despite the enhancement of $1/T_1$, $\chi$ is suppressed from the Curie-Weiss behavior, $\chi_{\rm CW} = 1/(4T-J)$,
which is obtained by the standard mean-field approximation in the original spin representation. 
These $T$ dependences of $1/T_1$, $\chi$, and $\langle S^{p}_j S^{p}_k \rangle$ below $T_{\rm H}$ are highly unusual; in conventional quantum magnets, the dynamical spin correlations grow with the static ones. 
The dichotomy between the static and dynamical correlations is a clear signature of fractionalization of quantum spins~\cite{supp}.

In summary, we have presented a comprehensive set of theoretical results for dynamical and static spin correlations, which evince fluctuating fractionalized spins in the Kitaev QSLs. 
The results are unveiled by using the Majorana CDMFT+CTQMC method developed in the current study.  
Experimentally, an unusual inelastic response, similar to our results of $S(\textbf{q},\omega)$, was observed in the recent neutron scattering experiment for a Kitaev candidate, $\alpha$-RuCl$_3$~\cite{Banerjee2016}. 
Also, a similar peak in $1/T_1$ to our results was observed for another candidate, Li$_2$RhO$_3$~\cite{Khuntia_preprint}. 
The deviation of $\chi$ from $\chi_{\rm CW}$ was already reported in many materials~\cite{Singh2010,Singh2012,Plumb2014}. 
Obviously, further systematic studies for the Kitaev candidate materials are highly desired to test our predictions. 
Our results will stimulate the rapidly-evolving ``pincer attack" by theory and experiment for the long standing issue---the identification of fractionalized spins in QSLs. 

\begin{acknowledgments}
The authors thank M. Imada, Y. Kato, M. Udagawa, and Y. Yamaji for fruitful discussions.
This research was supported by KAKENHI (No.~24340076 and 15K13533), the Strategic Programs for Innovative Research (SPIRE), MEXT, and the Computational Materials Science Initiative (CMSI), Japan.
\end{acknowledgments}

\appendix                                                           %
\vspace{15pt}                                                       %
\begin{center}                                                      %
  {\bf ---Supplemental Material---}                                 %
\end{center}                                                        %
\setcounter{figure}{0}
\setcounter{equation}{0}
\setcounter{table}{0}
\renewcommand{\thefigure}{S\arabic{figure}}
\renewcommand{\theequation}{S\arabic{equation}}
\renewcommand{\thetable}{S\Roman{table}}

\section*{\large Cluster size dependence}

The CDMFT is an approximation which replaces the infinite-size system by a finite-size cluster embedded in a bath [see Fig.~1(a) in the main text]. 
It becomes exact when the cluster size is increased to infinity.
Thus, it is crucial how the results converge to the thermodynamic limit as a function of the cluster size. 

\begin{figure}[htb]
\begin{center}
\includegraphics[width=0.9\columnwidth,clip]{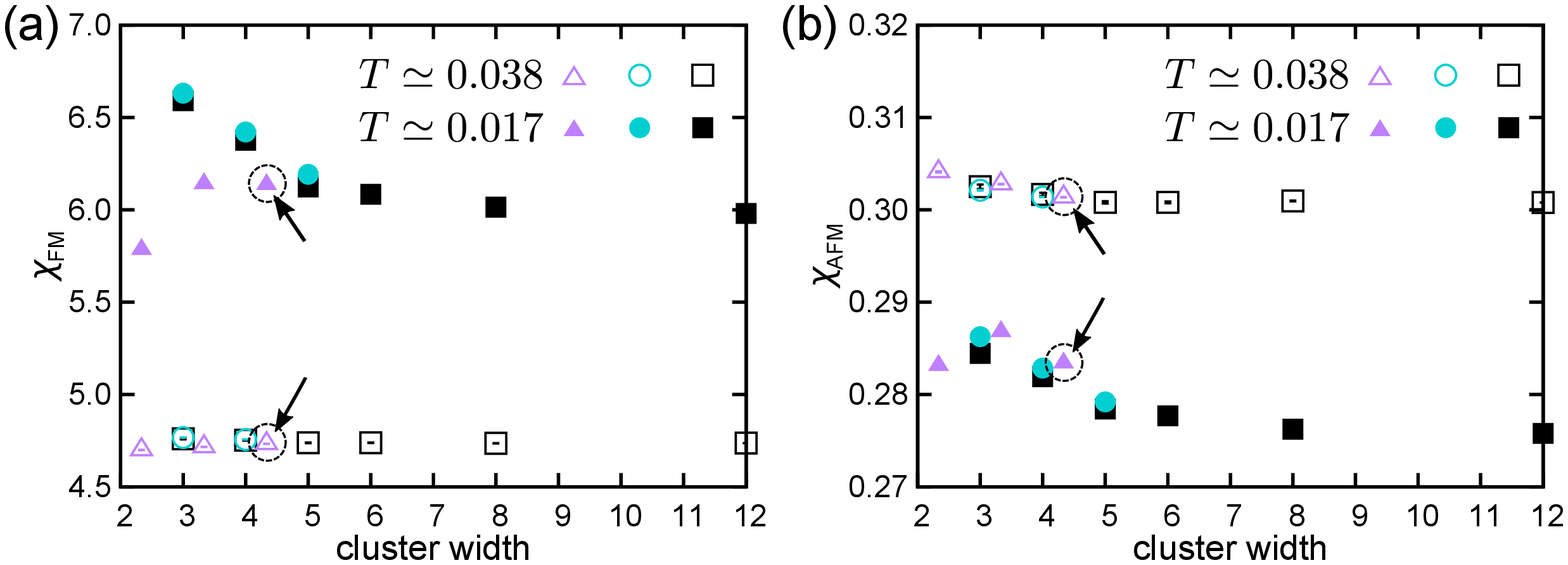}
\caption{
Cluster size dependence of the magnetic susceptibility for (a) the FM and (b) AFM cases. 
The different symbols represent the different series of the clusters: 
triangles, circles, and squares correspond to Fig.~\ref{fig:ClusterType}(a), \ref{fig:ClusterType}(b), and \ref{fig:ClusterType}(c), respectively. 
The data in the dashed circles with arrows indicate the results for the 26-site cluster used in the calculations in the main text. 
The definition of the cluster width is described in the caption of Fig.~\ref{fig:ClusterType}. 
}
\label{fig:size_dep}
\end{center}
\end{figure}

Figure~\ref{fig:size_dep} shows the cluster size dependence of the magnetic susceptibility obtained by the CDMFT+CTQMC method. 
The data are plotted as functions of the cluster width in the $xy$ direction, not the total number of lattice sites included in the cluster. 
This is because the width in the $xy$ direction is rather relevant compared to that in the $z$ direction in the present CDMFT, presumably due to the Majorana representation based on the Jordan-Wigner transformation along the $xy$ chain. 
Other physical quantities calculated in the main text behave in a similar manner. 

\begin{figure}[htb]
\begin{center}
\includegraphics[width=0.9\columnwidth,clip]{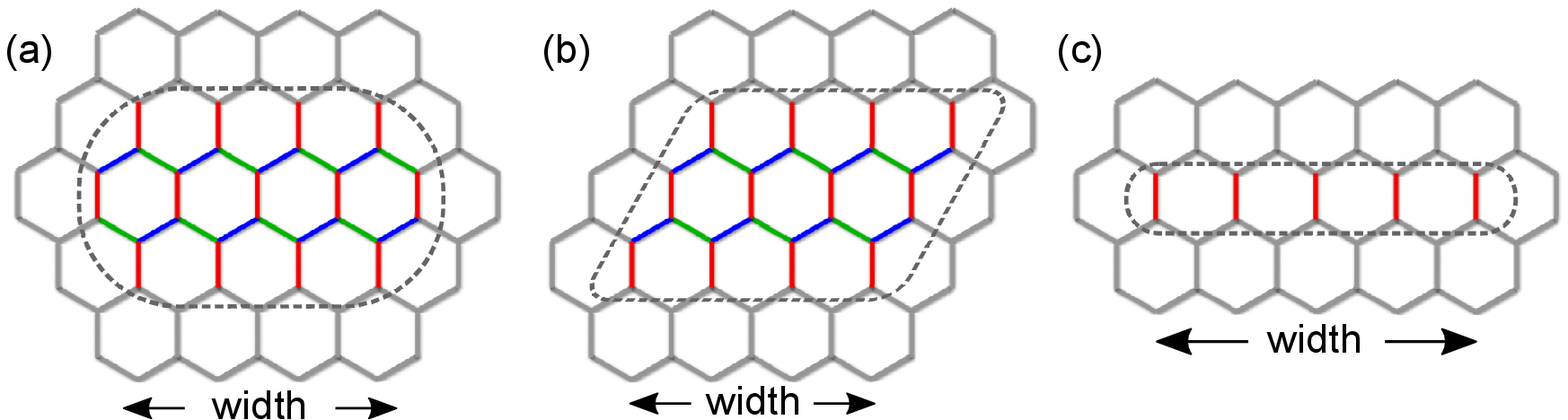}
\caption{Schematic picture of three series of the clusters used in Fig.~\ref{fig:size_dep}. 
In each series of clusters, the cluster size is varied by the cluster width in the $xy$ direction, while keeping the width in the $z$ direction. 
The cluster width in the $xy$ direction is given by the average number of the $z$-bonds (indicated by the red lines in the figures), which is used in the plots in Fig.~\ref{fig:size_dep}. 
In these examples, the cluster width is (a) $13/3 \simeq 4.3$, (b) 4, and  (c) 5.
}
\label{fig:ClusterType}
\end{center}
\end{figure}

As shown in Fig.~\ref{fig:size_dep}, for both FM and AFM cases, the results show good convergence when increasing the cluster size. 
In fact, the convergence is very quick, except for the low-$T$ region in the vicinity of the artificial transition temperature $T_c\simeq 0.014$, which is close to $T_{\rm L}\simeq 0.012$ [see Fig.~1(b) in the main text]. 
For instance, the data at $T\simeq 0.038$, which is sufficiently high compared to $T_c$, are almost unchanged while increasing the cluster width larger than $\sim 4$ in all the different series of the clusters. 
On the other hand, while lowering temperature and approaching $T_c$, the cluster size dependence becomes substantial, as shown in the data at $T\simeq 0.017$ in the figure. 
Nevertheless, the data for the 26-site cluster, which is used in the main text [the width is $\sim 4.3$ in the series of Fig.~\ref{fig:ClusterType}(a)], give sufficiently converged results: the remnant relative errors are $\lesssim 3$\% for both FM and AFM cases. 
Note that the remnant errors become discernible only in the very vicinity of $T_c$. 
From these observations, we confirmed that our data in the wide range of $T$ above $T_c\simeq 0.014$ are quantitatively correct and well reproduce the behaviors expected in the thermodynamic limit.

\section*{\large CDMFT+CTQMC results for $T<T_{\rm L}$}

\begin{figure}[htb]
\begin{center}
\includegraphics[width=0.9\columnwidth,clip]{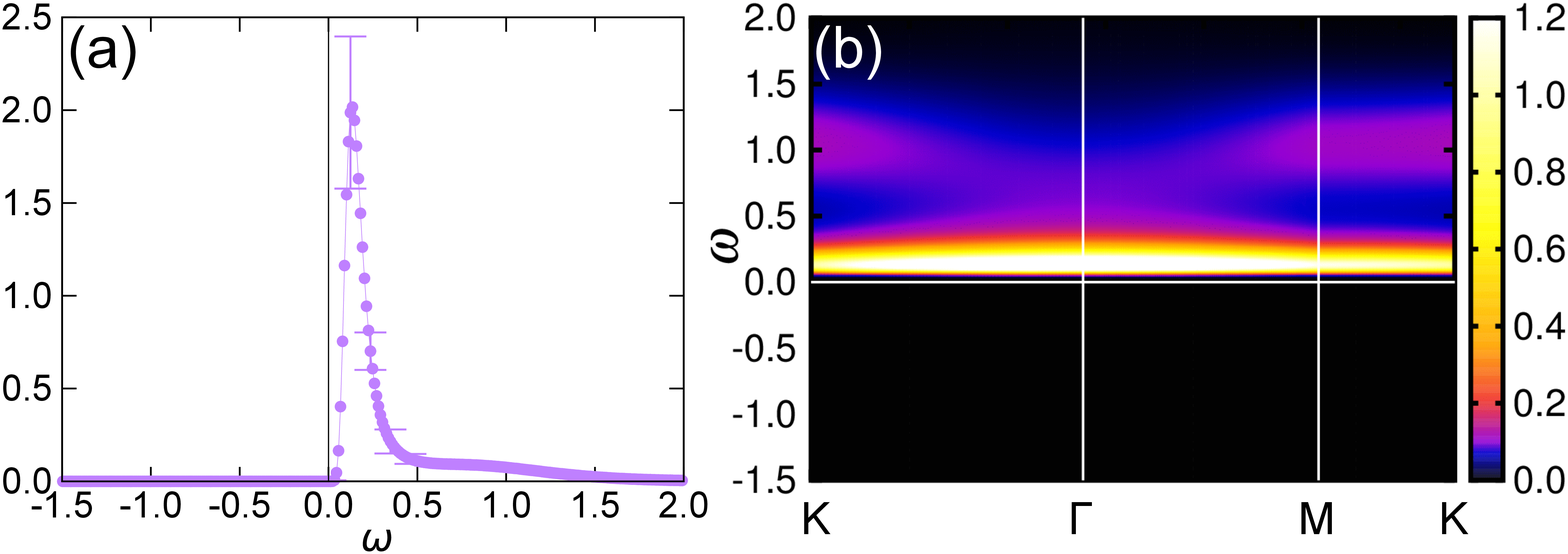}
\caption{
(a) $S(\Gamma,\omega)$ and (b) $S({\bf q},\omega)$ at $T=0.00825$ for the FM case. 
In (a), the errorbars are shown for every ten data along the $\omega$ axis.
}
\label{fig:lowestT}
\end{center}
\end{figure}

As mentioned in the main text, the CDMFT calculation exhibits a phase transition by ordering of $\langle \eta_r \rangle$ at $T \simeq 0.014$ because of the mean-field nature of CDMFT. 
Below the critical temperature, $\langle \eta_r \rangle$ becomes almost $1$, namely, the state is almost similar to the flux-free ground state. 
In the ground state, $S({\bf q},\omega)$ shows a small gap $\simeq 0.065J$ due to the gapped flux excitation~\cite{Knolle2014}. 
In Fig.~\ref{fig:lowestT}, we show the CDMFT+CTQMC results for the dynamical spin structure factor at $T=0.00825$, which is well below the critical temperature as well as $T_{\rm L}$. 
The result exhibits the flux gap, consistent with the previous results at $T=0$. 
This further supports the validity of our CDMFT+CTQMC calculations. 
In Figs.~2 and 3 in the main text, the flux gap is smeared out and not clearly visible as the fluxes are excited by thermal fluctuations above $T_{\rm L}$. 

\section*{\large Plots of Fig.~4 in the $T$-linear scale}

\begin{figure}[htb]
\begin{center}
\includegraphics[width=0.9\columnwidth,clip]{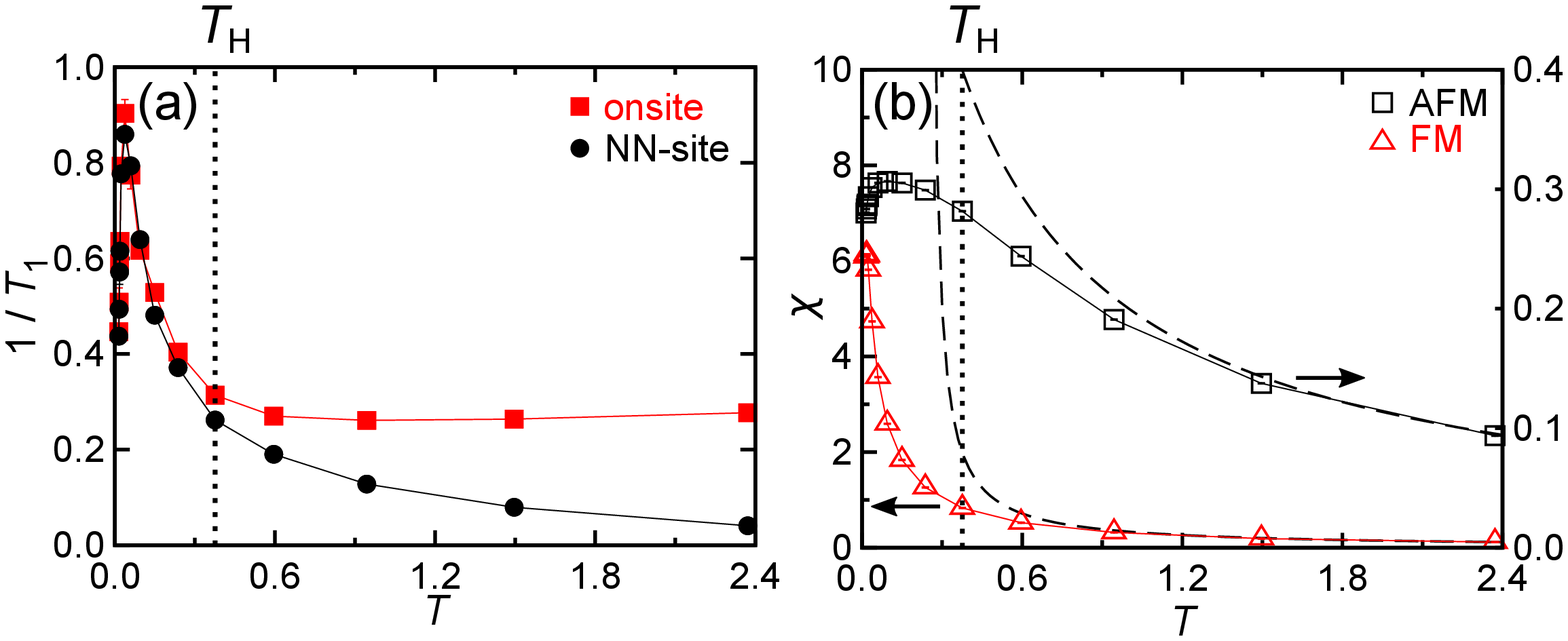}
\caption{
The same plots as Fig.~4 in the main text in the $T$-linear scale.
    }
\label{fig:T-linear}
\end{center}
\end{figure}

In Fig.~4 in the main text, we show the NMR relaxation rate $1/T_1$ and the magnetic susceptibility $\chi$ as functions of $\ln T$. 
For reference, we present them in the $T$-linear scale in Fig.~\ref{fig:T-linear}. 

                               %


\begin{thebibliography}{999}
\bibitem{note2} The difference between the two components corresponds to $S(\Gamma,\omega=0)$ in the AFM case [see Fig.~\ref{fig:fig3}(d)]. 
It becomes small after the growth of spin correlations below $T_{\rm H}$, as it is proportional to the probability of aligning two spins on a $z$ bond parallel to the $z$ direction without an energy cost.
\end{thebibliography}
\end{document}